\documentclass[a4paper]{article}
\usepackage[T1]{fontenc}
\usepackage{lmodern} 

\usepackage[a4paper, margin=1in]{geometry}

\usepackage{authblk}
\usepackage{setspace}
\usepackage{graphicx}
\graphicspath{{./figures/}}
\usepackage{subcaption}
\usepackage{amsmath, amssymb, mathrsfs}
\usepackage{float}
\usepackage{balance}
\usepackage{enumitem, array}
\usepackage{multirow, booktabs}
\usepackage{tabularx}
\usepackage{comment}
\usepackage{booktabs}
\usepackage{adjustbox} 
\usepackage{pdflscape}
\usepackage{url}
\usepackage{geometry}
\usepackage{chemformula} 
\usepackage{siunitx}     
\usepackage[version=4]{mhchem}
\usepackage{gensymb} 

\usepackage[colorlinks=true, citecolor=blue, linkcolor=blue, urlcolor=blue]{hyperref}

\usepackage[
    style=numeric-comp, 
    sorting=none,       
    backend=biber,      
    defernumbers=true,
    maxcitenames=2,
    giveninits=true
]{biblatex}
\usepackage{lineno}

\usepackage[labelfont=bf]{caption}

\title{Carbon mineralization in CO$_2$-seawater-basalt systems: Reactive transport dynamics and vesicular pore architecture controls}

\author[1*]{Mohammad Nooraiepour}
\author[2]{Mohammad Masoudi}
\author[1]{Helge Hellevang}

\affil[1]{\small {Environmental Geosciences, Department of Geosciences, University of Oslo, 0316 Blindern, Oslo, Norway.}}
\affil[2]{\small {SINTEF Industry, Applied Geoscience Department, 7465 Trondheim, Norway}}
\affil[*]{\small{Corresponding author: mohammad.nooraiepour@geo.uio.no}}

\date{}

\begin{document}
\maketitle

\begin{abstract}
\noindent Carbon mineralization in basaltic rocks may offer rapid, permanent \ce{CO2} storage, yet fundamental controls on reactive transport and precipitation patterns remain poorly understood. This study integrates flow-through experiments at 80\degree C using \ce{CO2}-acidified seawater with geochemical simulation and multi-scale pore imaging to elucidate mineralization dynamics in basaltic glass. Results reveal that carbonate precipitation is nucleation-controlled and stochastic rather than growth-controlled and deterministic, with isolated accumulations forming randomly despite continuous supersaturation. Residence time exerts primary control: reducing flow rate from 0.05 to 0.005\,mL/min proved necessary for visible precipitation. Post-experiment analyses identified calcium carbonate and smectite phases. Multi-scale characterization of three basalt facies revealed that connected porosity fractions (1.3--32\%) differ significantly from total porosity (18--42\%), demonstrating that network topology controls permeability. Micro-CT analysis revealed that pore coordination numbers in basalts (modal = 2) were notably lower than those in reservoir sandstones, creating serial flow paths that are vulnerable to catastrophic permeability loss from modest precipitation. Precipitation-induced clogging scenarios were proposed, where distributed small precipitates cause more severe permeability degradation than large accumulations. The use of seawater complicates geochemistry and reduces mineralization efficiency compared to freshwater. Findings emphasize the need for probabilistic reactive transport modeling frameworks and realistic pore topologies, which are fundamentally different from conventional CCS operations.
\end{abstract}

\section{Introduction}
The escalating climate crisis, driven by human-made carbon dioxide (\ce{CO2}) emissions, underscores the urgent need for scalable solutions to mitigate the adverse effects of climate change. Atmospheric \ce{CO2} concentrations have surged to unprecedented levels, driving rising global temperatures, ocean acidification, and extreme weather events \cite{kikstra2022ipcc, zhang2022globalization}. To limit global warming to 1.5--2\,\textdegree{}C above pre-industrial levels, as outlined in the Paris Agreement \cite{schurer2017importance}, it is imperative to not only reduce \ce{CO2} emissions but also deploy technologies for capturing and storing \ce{CO2} at scale. Carbon capture and storage (CCS) and geological carbon storage (GCS) have emerged as a viable strategy to sequester large volumes of \ce{CO2} safely and permanently \cite{ringrose2021storage, Hellevang2015, nooraiepour2025norwegian}.

Saline aquifers are the primary choice for geological \ce{CO2} sequestration due to their large storage capacity, widespread availability, and proximity to emission sources \cite{ringrose2021storage, izadpanahi2024review,nooraiepour2025geologicalco2storageassessment}. However, long-term storage security in saline aquifers relies heavily on caprock integrity to prevent buoyant \ce{CO2} migration to the surface \cite{song2023geomechanical, gholami2021leakage}, as these systems primarily target structural/stratigraphic and residual trapping mechanisms. In contrast, GCS in mafic and ultramafic rocks offers a promising alternative with the added benefit of \ce{CO2} mineralization, achieving rapid mineral trapping \cite{oelkers2023carbon, sandalow2021carbon, kelemen2019overview, nisbet2024carbon}. Basaltic rocks, rich in divalent cations such as calcium (\ce{Ca^2+}), magnesium (\ce{Mg^2+}), and iron (\ce{Fe^2+}) in silicate minerals, react with \ce{CO2} to form stable carbonate minerals. The mineral trapping mechanism represents the most secure form of carbon storage.

The potential of mafic and ultramafic formations for carbon mineralization has been demonstrated through several pioneering projects worldwide, including the CarbFix project in Iceland \cite{matter2016rapid, oelkers2023carbon}, the Wallula pilot project in Washington, USA \cite{white2020quantification, mcgrail2017wallula}, and emerging initiatives such as \ce{CO2}Lock \cite{mosavat2024brucite}, CarbonStone \cite{teboul2024carbonstone}, and 44.01 \cite{rassenfoss2023mountains}. The CarbFix project leverages \ce{CO2} dissolution in freshwater to form carbonic acid, which reacts with basaltic rocks to precipitate carbonate minerals, with a reported 95\% mineralization rate within two years at 20--50\,\textdegree{}C \cite{matter2016rapid, oelkers2023carbon}. However, this approach requires substantial freshwater resources---approximately 22 metric tons per ton of \ce{CO2} at 30\,bar and 20\,\textdegree{}C---raising scalability concerns and potential conflicts with water resources. In contrast, the Wallula project directly injects supercritical \ce{CO2} (sc\ce{CO2}) into basaltic formations, relying on caprock integrity for containment, with hydrological modeling suggesting up to 60\% mineralization within two years \cite{white2020quantification, mcgrail2017wallula, nisbet2024carbon}.

These projects demonstrate the feasibility of in-situ mineral carbonation in mafic and ultramafic rocks under localized conditions. However, global scalability requires expanding field sites and addressing key challenges: optimizing \ce{CO2} injection rates, ensuring consistent long-term fluid movement within storage reservoirs, and enhancing dissolution--precipitation rates. Despite the successes of pilot projects, deriving detailed hydrological, chemical, and mechanical parameters from field observations alone remains challenging. Critical gaps persist in scaling this technology globally. First, reported mineralization rates must be sustained at megaton-scale injections, requiring a deeper understanding of geochemical and reactive transport processes under field conditions. Second, seawater or in-situ saline water must replace freshwater to enhance applicability, particularly for offshore basalt deposits. These offshore basaltic formations offer immense theoretical storage potential---up to 100,000 gigatonnes of \ce{CO2}, exceeding 2,000 times current annual global emissions. Additional critical questions requiring urgent attention, include: (i) evaluating sc\ce{CO2} injection regarding reactive transport and containment integrity, which requires secure caprocks and extended retention times for complete mineralization; (ii) determining realistic timescales for mineral dissolution and precipitation processes to provide industrial-scale timeframes for mineral trapping; and (iii) identifying reactive surface areas and reaction rates in fractures (where advection dominates) versus pores and matrices (where diffusion prevails) to enable stochastic simulation of reactive flow scenarios.

Despite numerous reviews on carbon mineralization \cite{nisbet2024carbon, raza2022carbon, cao2024review, kim2023review, otabir2025geochemical, rasool2023reactivity, romanov2015mineralization, lu2024knowledge, lu2024carbon}, the geochemical interactions and reactive transport dynamics of \ce{CO2}-rich seawater in basalts remain poorly understood. To address knowledge gaps regarding seawater use as the aqueous phase under reactive flow conditions, this study integrates laboratory experiments with geochemical fluid-rock interaction simulations to assess \ce{CO2} mineralization dynamics. We focus on three key aspects: (1) the potential for rapid \ce{CO2} mineralization using seawater under reactive transport conditions, (2) the effect of residence time on mineralization through advection velocity variations, and (3) pore space architecture and characteristics of vesicular basalt reservoirs.

\section{Challenges of carbon mineralization in basalts using seawater}
Carbon mineralization in basaltic rocks using \ce{CO2}-charged seawater presents significant complexities due to the competing reactions that occur under diverse thermodynamic and geochemical conditions. The substitution of seawater for freshwater introduces additional challenges affecting carbonate nucleation and growth, creating substantial uncertainties in reaction pathways and efficiency due to heterogeneous mineral-fluid interactions.

Experimental studies reveal that \ce{CO2}-charged seawater interactions with basaltic rocks are highly temperature- and composition-dependent. Carbon mineralization efficiency is maximized at 90--150\,\textdegree{}C, where silicate dissolution and carbonate precipitation rates are optimally balanced. Lower temperatures (<50\,\textdegree{}C) result in slower reaction kinetics, while temperatures above 200\,\textdegree{}C cause retrograde solubility and formation of competing phases such as zeolites, decreasing yields. Rosenbauer et al. \cite{rosenbauer2012carbon} demonstrated ferroan magnesite formation at 100\,\textdegree{}C achieving 8--26\% \ce{CO2} mineralization, while Shibuya et al. \cite{shibuya2013reactions} observed calcite as the dominant precipitate at 250--350\,\textdegree{}C, reducing dissolved \ce{CO2} concentrations by 75--100\%.

The high \ce{Mg^2+} content in seawater competes with \ce{Ca^2+} for carbonate incorporation, often forming mixed Ca-Mg carbonates with lower stability than pure calcite or magnesite \cite{Xu2013, Loste2003, Rodriguez-Blanco2020}. The presence of sulfate catalyzes the precipitation of anhydrite and gypsum (\ce{CaSO4}), thereby reducing porosity and sequestering essential \ce{Ca^2+} ions. Rigopoulos et al. \cite{rigopoulos2018carbon} demonstrated that removing sulfate from artificial seawater minimized anhydrite interference, increasing carbonate yield by 30\%. However, replicating these conditions in natural seawater systems adds operational complexity, particularly for offshore implementations.

Kinetic limitations arise from the formation of metastable byproducts. Voigt et al. \cite{voigt2021experimental} showed that high \ce{CO2} partial pressure (16\,bar) favors magnesite over Mg-rich clays in submarine basalts, mineralizing 20\% of \ce{CO2} in five months. At lower pressures, clay formation consumes \ce{Mg^2+} and stalls carbonate growth. Wolff-Boenisch and Galeczka \cite{wolff2018flow} demonstrated that induced Ca/Mg-carbonate precipitation requires artificially elevated carbonate saturation, which is rarely achieved in natural reservoirs. Without engineered supersaturation, secondary minerals like clays and zeolites sequester crucial cations, making nucleation less energetically favorable. Without engineered supersaturation, secondary minerals like clays and zeolites sequester crucial cations, reducing free \ce{Mg^2+} (and \ce{Ca^2+}) concentrations and thus lowering carbonate supersaturation ($\Omega$). This increases the nucleation energy barrier for magnesite, rendering spontaneous carbonate formation kinetically unfavorable.

The efficiency of seawater-based systems depends critically on the optimization of temperature and pH. Lower temperatures and acidic conditions inhibit carbonate precipitation or favor non-carbonate phases such as smectites \cite{hellevang2017experimental}. Robust carbonate formation occurs at elevated temperatures and near-neutral pH (7.6--8.7), emphasizing the importance of optimizing reaction conditions.

While numerous batch-type laboratory studies \cite{brady1997seafloor, wolff2011dissolution, marieni2020experimental, stavropoulou2024impact, kopf2024initial, voigt2021experimental} have advanced understanding of basalt-seawater-\ce{CO2} interactions, translating these findings to field-scale applications remains challenging. These investigations underscore the need for further geochemical and reactive transport studies, complemented by the characterization of secondary minerals, particularly in submarine environments. Extensive research is necessary to optimize reaction conditions for seawater-based carbon mineralization in basaltic systems, particularly at lower temperatures and near-neutral pH, in order to elucidate the complex balance that governs successful outcomes..

\section{Materials and Methods}

\subsection{Flow-through column reactor experiments}
Flow-through reactive transport experiments were conducted using a custom-designed glass tube reactor (40\,cm length, 14\,mm inner diameter). The reactor was filled and carefully packed with two distinct zones: crushed calcium carbonate grains in the initial 4\,cm section to provide dissolved \ce{Ca^2+} and pH buffering after reaction with injected \ce{CO2}-acidified seawater, and basaltic glass occupying the remaining 36\,cm as the primary substrate for carbonate nucleation and growth. 

Basaltic glass from Stapafell Mountain, Reykjanes Peninsula, Iceland, served as the primary reactive substrate for the column experiments. This tholeiitic basalt, with composition comparable to mid-ocean ridge basalt (MORB) (Table~\ref{tab:basalt_composition}), exhibits a divalent cation oxide content (\ce{CaO} + \ce{MgO} + \ce{Fe2O3} = 23.8 wt\%) optimal for carbon mineralization.

The crushed calcium carbonate and basaltic grains were ultrasonically cleaned in Milli-Q ultrapure deionized water and subsequently dried at ambient temperature before being added to the reactor. The basaltic glass, sourced from Stapafell (Reykjanes Peninsula, Iceland), was selected for its rapid reactivity, homogeneous chemical composition, and dark coloration, which facilitates the identification of secondary carbonate phases.

Seawater (approximately 1\,L) was collected from Oslo Fjord, filtered through 0.45\,\textmu{}m Millipore\textsuperscript{\textregistered} filters, and used as the base fluid due to its chemical similarity to average North Sea seawater. The seawater was acidified by \ce{CO2} dissolution under 4\,MPa pressure in a pressurized fluid accumulator housed within a forced convection benchtop oven. The equilibrated \ce{CO2}-charged seawater was injected using a dual-piston ISCO pump at two controlled flow rates: 0.05 and 0.005\,mL/min, enabling evaluation of residence time effects on mineral precipitation kinetics. This flow rate variation provided residence times that allowed assessment of advection velocity impacts on carbonate formation.

The column had a total volume of $61.6$\,mL and a pore volume of approximately $25$\,mL. At the higher flow rate of $0.05$\,mL/min, this corresponds to $\sim$ {1 PV} exchanged every $\sim$8.2\,h, with a Darcy velocity of $5.4 \times 10^{-6}$\,m/s ($0.032$\,cm/min) and an interstitial (pore) velocity of $1.4 \times 10^{-5}$\,m/s ($0.081$\,cm/min). At the lower flow rate of $0.005$\,mL/min, $\sim$ {1 PV} was exchanged every $\sim$82\,h, with a Darcy velocity of $5.4 \times 10^{-7}$\,m/s ($0.0032$\,cm/min) and a pore velocity of $1.4 \times 10^{-6}$\,m/s ($0.0081$\,cm/min).

Experiments were conducted at 80\,\textdegree{}C with injection pressure of approximately 1\,MPa (automatically controlled by injection pumps) and atmospheric \ce{CO2} pressure at the outlet for 30\,days. Fluid samples were collected at discrete intervals (every 3\,days) from the column inlet and continuously monitored at the outlet. The solution pH was measured immediately after sampling using a calibrated benchtop pH meter (Metrohm 780 pH/ion meter) to track the geochemical evolution during the experiments. Detailed descriptions of the fluid injection system are provided in our prior studies \cite{nooraiepour2018effect,naseryan2019relative}.

\subsection{Sample characterization}
The mineralogical and elemental composition was identified and quantified using X-ray diffraction (XRD) and X-ray fluorescence (XRF), respectively, following the methodology described in our previous studies \cite{nooraiepour2017experimental,nooraiepour2017compaction}.

Scanning electron microscopy (SEM) was used to examine the surface structure and mineral growth on the substrates. Energy-dispersive X-ray spectroscopy (EDS) enabled chemical analysis and elemental mapping to identify mineral phases and their spatial distribution. A variable pressure Hitachi SU5000 FE-SEM (Schottky FEG) with a Dual Bruker XFlash system was employed for imaging and spectroscopy. Samples were coated with a thin carbon layer to enhance image quality, improve chemical analysis accuracy, and prevent surface charging. Further details are given in \cite{nooraiepour2021probabilistic}.

For pore space characterization, three distinct Icelandic basalt facies with varying degrees of vesicularity were analyzed, with two specimens per facies to capture heterogeneities and the range of variations. High-resolution X-ray microcomputed tomography (micro-CT) was used to investigate the pore architecture of vesicular basalts, providing three-dimensional insights into porosity and pore connectivity, which are crucial for understanding fluid flow and reactive transport processes. Image acquisition, processing, and segmentation followed the protocol described in our previous works \cite{nooraiepour2025potential,nooraiepour2025three}. Pore network modeling (PNM) was performed on segmented micro-CT datasets using OpenPNM \cite{gostick2016openpnm}. The extracted three-dimensional pore networks provided quantitative characterization of vertex (pore body) and edge (throat) properties, including connectivity, volume, diameter, and cross-sectional area distributions.

\subsection{Geochemical modeling}
Geochemical modeling was performed using PHREEQC v3 \cite{parkhurst2013description} to simulate aqueous geochemical processes, including advective flow and dispersion, solute saturation states, pH evolution, \ce{CO2} pressure dynamics, and reaction progress. For calculations, the standard state is assumed to be the unit activity for pure minerals and \ce{H2O} at specified temperature and pressure conditions, with a hypothetical one-molal solution referenced to infinite dilution for all aqueous species.

The CarbFix thermodynamic database \cite{voigt2018evaluation,} was used for all calculations, which in turn is constructed from the core10.dat database derived from the llnl.dat database distributed with PHREEQC. This database selection ensures accurate thermodynamic predictions of mineral dissolution and precipitation reactions under the specific temperature, pressure, and compositional conditions of this study. The CarbFix database has been validated explicitly for basalt-\ce{CO2}-water interactions, making it particularly suitable for modeling carbonate mineralization processes in mafic rock systems. The geochemical models served dual purposes: (1) interpreting experimental observations through thermodynamic equilibrium calculations and kinetic assessments, and (2) predicting long-term \ce{CO2} mineralization behavior in basaltic rocks under varying reactive transport conditions.

\section{Results and Discussion}
\subsection{Basaltic glass composition and characterization}
The basaltic glass used in our study was sourced from Stapafell Mountain on the Reykjanes Peninsula, southwest Iceland. This material exhibits a tholeiitic composition, characterized by high silica (\ce{SiO2}) and low sodium (\ce{Na2O}) content, with a fine-grained extrusive igneous texture. Prior to the experiments, the basaltic glass surface appeared smooth, showing no evidence of secondary mineral growth, though some cavities were observed. Secondary minerals formed only post-experimentation. 

XRD and XRF analyses confirmed that the composition of the Stapafell basaltic glass aligns closely with values reported in the literature \cite{oelkers2001mechanism,viennet2017dioctahedral,stockmann2011carbonate,gysi2011co2}. The major element composition, detailed in Table~\ref{tab:basalt_composition}, is comparable to the mid-ocean ridge basalt (MORB) \cite{ronov1976new,staudigel1998geochemical,oelkers2001mechanism}. This basaltic material has been extensively characterized and utilized in prior kinetic and experimental studies \cite{oelkers2001mechanism,gislason2003mechanism,hellevang2017experimental,flaathen2010effect,stockmann2011carbonate,hellevang2019kinetic,gysi2011co2,galeczka2014experimental,voigt2021experimental,}.

\begin{table}[h]
\footnotesize
\centering
\caption{Chemical composition (wt\%) of Stapafell basaltic glass, determined by XRF analysis.}
\label{tab:basalt_composition}
\begin{tabular}{@{}l *{15}{S[table-format=2.2]}@{}}
\toprule
& {\ce{SiO2}} & {\ce{TiO2}} & {\ce{Al2O3}} & {\ce{Fe2O3}} & {\ce{MnO}} & {\ce{MgO}} & {\ce{CaO}} & {\ce{Na2O}} & {\ce{K2O}} & {\ce{P2O5}} & {\ce{Cr2O3}} & {\ce{ZnO}} & {\ce{SO3}} \\
\midrule
wt\% & 47.282 & 1.487 & 14.657 & 12.148 & 0.213 & 10.079 & 11.438 & 1.756 & 0.273 & 0.206 & 0.109 & 0.013 & 0.111 \\
\bottomrule
\end{tabular}
{\footnotesize \parbox{\linewidth}{\raggedright \ce{Fe2O3} represents total iron including \ce{Fe^{2+}} and \ce{Fe^{3+}}}}
\end{table}

\begin{figure}[h!]
    \centering
    \includegraphics[width=0.9\textwidth]{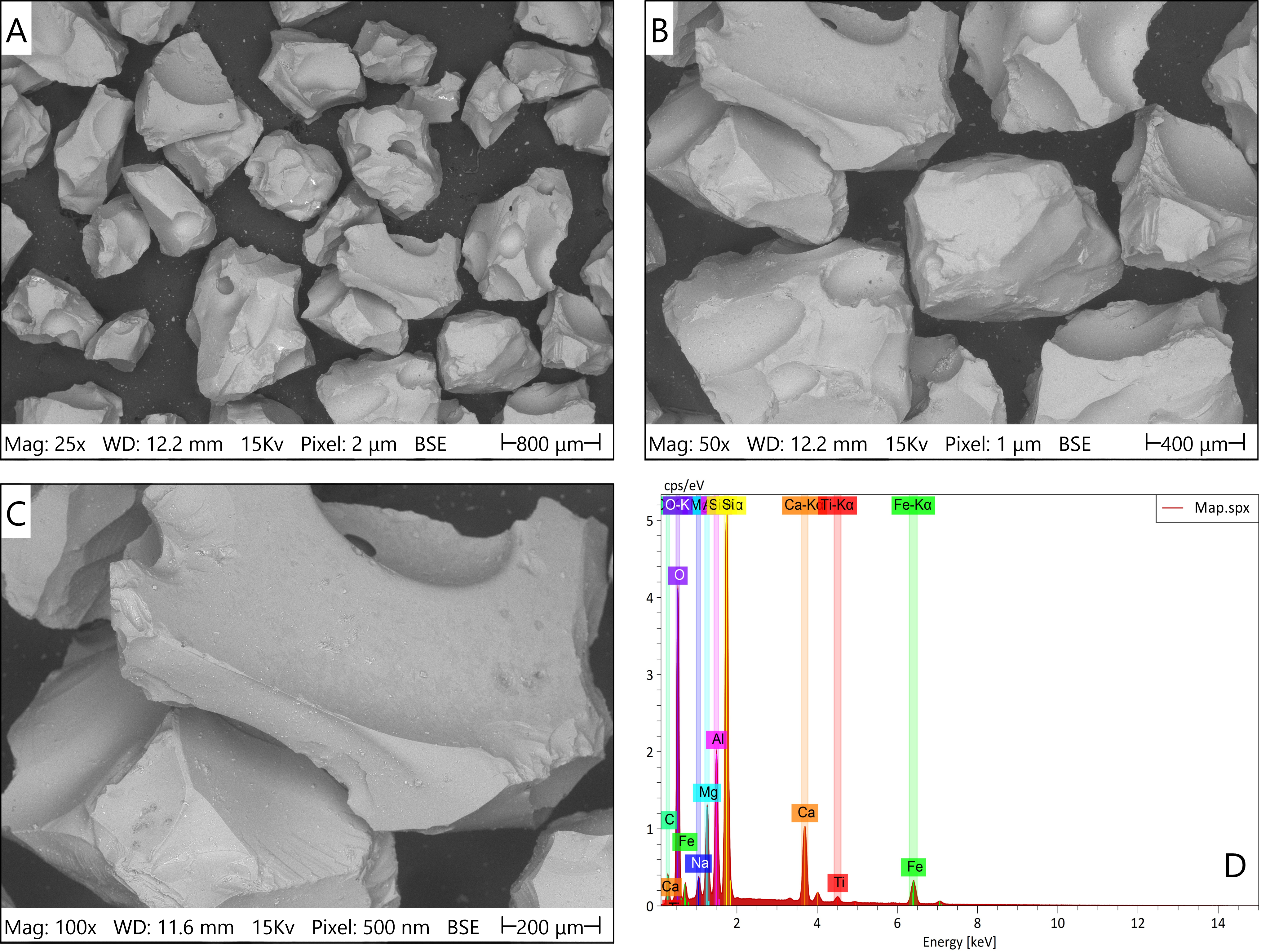}
    \caption{Multi-scale SEM characterization of unreacted Stapafell basaltic glass grains used in flow-through experiments. (a) Low-magnification overview (25$\times$, pixel size = \SI{2}{\micro\meter}) showing grain size distribution and morphology. (b) Intermediate magnification (50$\times$, pixel size = \SI{1}{\micro\meter}) revealing smooth glassy surfaces characteristic of rapidly cooled volcanic material. (c) High-magnification image (100$\times$, pixel size = \SI{0.5}{\micro\meter}) displaying surface features including natural cavities, vesicular structures, and conchoidal fracture patterns, with no evidence of pre-existing secondary mineral phases. (d) Representative EDS spectrum from pristine grain surfaces, confirming the presence of major elements (Si, Al, Fe, Ca, Mg, O) consistent with tholeiitic basalt composition and demonstrating the availability of reactive divalent cations (Ca\textsuperscript{2+}, Mg\textsuperscript{2+}, Fe\textsuperscript{2+}) for \ce{CO2} mineralization reactions.}
    \label{fig:basalt_sem_pristine}
\end{figure}

A detailed microscopic examination of the unreacted basaltic glass grains was conducted to establish the baseline surface morphology and compositional characteristics prior to reactive transport experiments. Figure~\ref{fig:basalt_sem_pristine} presents multi-scale SEM characterization of the pristine basaltic glass substrate. Low-magnification imaging (25$\times$, pixel size = \SI{2}{\micro\meter}) reveals the overall grain morphology and size distribution of the crushed basaltic material (Fig.~\ref{fig:basalt_sem_pristine}a). At intermediate magnification (50$\times$, pixel size = \SI{1}{\micro\meter}), individual grain surfaces exhibit characteristically smooth, glassy textures with minimal surface roughness (Fig.~\ref{fig:basalt_sem_pristine}b). High-magnification imaging (100$\times$, pixel size = \SI{0.5}{\micro\meter}) provides a detailed visualization of surface features, including natural cavities, vesicular structures, and conchoidal fracture patterns typical of rapidly quenched volcanic glass (Fig.~\ref{fig:basalt_sem_pristine}c). These surface irregularities, while present, show no evidence of pre-existing secondary mineral phases or weathering products. The absence of surface alteration products confirms the suitability of this material as a pristine substrate for investigating \ce{CO2}-induced mineralization processes. Energy-dispersive X-ray spectroscopy (EDS) analysis of the unreacted grain surfaces (Fig.~\ref{fig:basalt_sem_pristine}d) confirms the presence of major constituent elements consistent with the bulk XRF composition presented in Table~\ref{tab:basalt_composition}. The EDS spectrum exhibits characteristic peaks corresponding to the tholeiitic basaltic composition, with prominent signals for Mg, Ca, and Fe, indicating the availability of reactive divalent cations essential for carbonate mineral precipitation.

The geochemical composition of the Stapafell basaltic glass is particularly well-suited for carbon mineralization. With a combined divalent cation oxide content (CaO + MgO + \ce{Fe2O3} = 23.8 wt\%) approaching the ~25 wt\% threshold of highly reactive basalts, this material exhibits optimal reactivity characteristics. The elevated CaO content (11.44 wt\%) is particularly important, as calcium represents the dominant cation for carbonate precipitation in CO\textsubscript{2} mineralization processes. Additionally, the substantial MgO content (10.08 wt\%) enhances the overall reactivity by supplying essential metallic divalent cations that facilitate rapid carbonate formation.

The amorphous nature of this basaltic glass confers substantial kinetic advantages over crystalline basaltic rocks, with dissolution rates markedly exceeding those of hydrothermally altered basalts \cite{wolff2006effect}. This enhanced reactivity results from two key factors: the metastable glassy matrix, which facilitates rapid initial dissolution, and the absence of passivating secondary mineral coatings that commonly form protective barriers in altered rocks.

\subsection{Reactive transport dynamics and carbonate mineralization}

One of the primary objectives of this study was to evaluate how advection velocity and fluid residence time influence carbonate mineral precipitation during \ce{CO2}-charged seawater interaction with basaltic glass under continuous flow conditions. Two controlled flow rates, differing by an order of magnitude (\SI{0.05}{mL/min} and \SI{0.005}{mL/min}), were employed to systematically assess the effect of residence time on mineralization kinetics and spatial precipitation patterns. These corresponded to Darcy velocities of \SI{5.4e-6}{m/s} and \SI{5.4e-7}{m/s}, respectively, yielding mean fluid residence times of approximately \SI{8.2}{h} and \SI{82}{h}.

Initial experiments conducted at the higher flow rate of \SI{0.05}{mL/min} for 30 days revealed no visible carbonate precipitates on basaltic glass surfaces, despite the continuous injection of \ce{CO2}-acidified seawater. Replicate experiments confirmed this absence of macroscopic carbonate formation, indicating that the observed phenomenon was not an experimental artifact. High-resolution SEM-EDS characterization of post-experiment substrates revealed only minor, early-stage pre-crystalline phases with elevated calcium (\ce{Ca}) content deposited on the primary basaltic glass substrate, detectable solely at high magnification. These findings indicated that the \SI{0.05}{mL/min} flow rate, corresponding to higher advection velocity and shorter fluid-rock contact time, was insufficient for significant carbonate nucleation and growth. The kinetic limitations at this flow rate suggest that carbonate supersaturation levels, while potentially achieved locally, did not persist long enough to overcome the nucleation energy barrier required for stable crystal formation.

Consequently, the flow rate was reduced by an order of magnitude to \SI{0.005}{mL/min}, increasing the residence time. This modification created more favorable thermodynamic and kinetic conditions for carbonate nucleation and growth by allowing extended fluid-rock interaction, progressive accumulation of dissolved divalent cations, and sustained supersaturation with respect to carbonate minerals. Under these conditions, visible carbonate precipitation was successfully achieved, enabling detailed characterization of mineralization processes and spatial distribution patterns.

\begin{figure}[h!]
    \centering
    \includegraphics[width=0.95\textwidth]{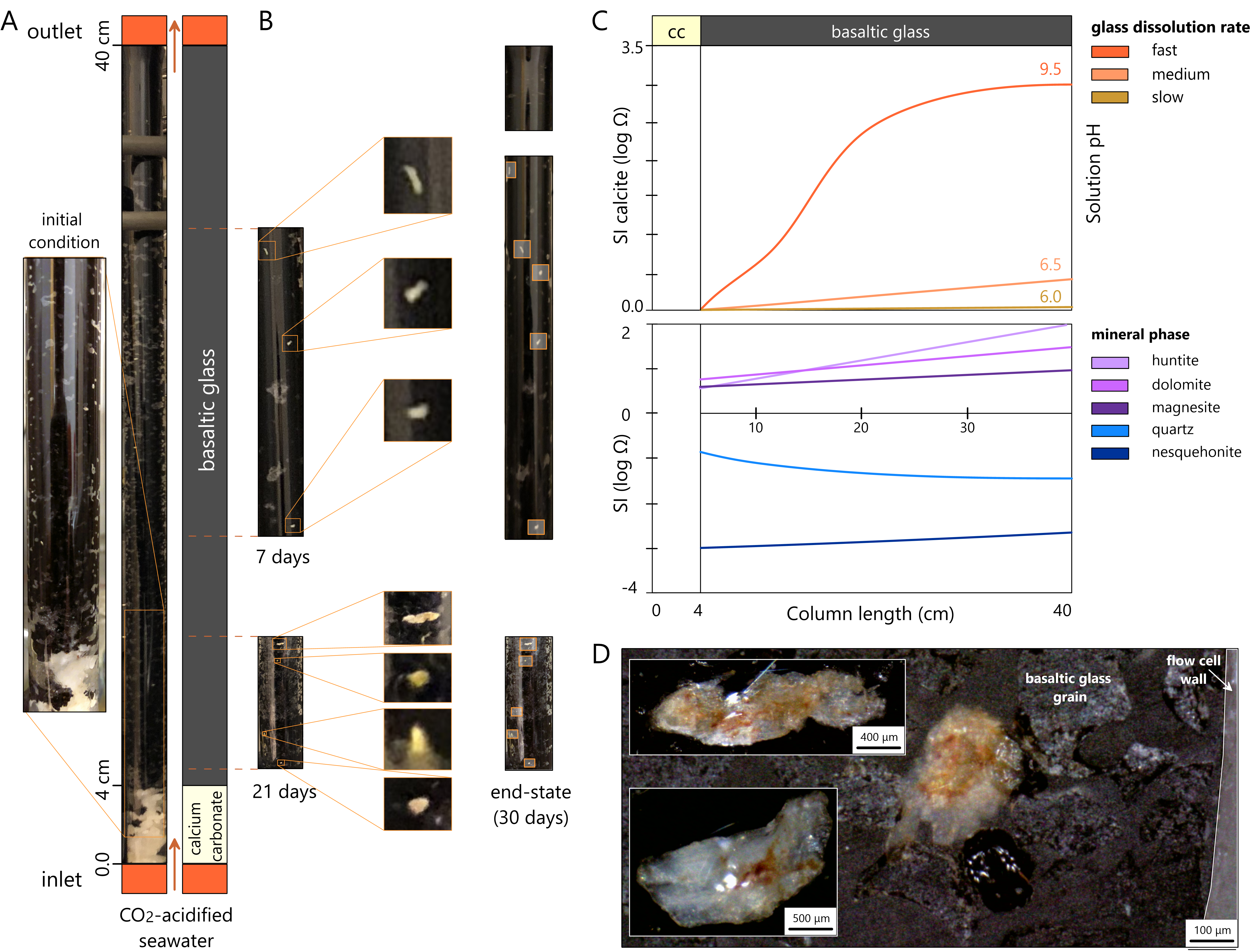}
    \caption{Reactive transport experiment results and geochemical modeling of \ce{CO2} mineralization in basaltic glass columns. (a) Experimental setup showing the column reactor configuration with an initial \SI{4}{cm} calcite section and \SI{36}{cm} basaltic glass section during \ce{CO2}-acidified seawater injection at \SI{0.005}{mL/min}. (b) Temporal evolution of carbonate mineral precipitation at 7, 21, and 30 days, demonstrating random spatial distribution of white carbonate accumulations along the basalt section, with preferential formation of larger precipitation patches in the latter half of the column. (c) PHREEQC v3 reactive transport simulation results: (top) calcite saturation index evolution for three basalt glass dissolution rate scenarios, fast (k = \SI{1.0e-8}{mol.m^{-2}.s^{-1}}), medium (k = \SI{1.0e-9}{mol.m^{-2}.s^{-1}}), and slow (k = \SI{1.0e-10}{mol.m^{-2}.s^{-1}}) with estimated outlet pH constraining actual rates near the medium case. (bottom) saturation indices of carbonate minerals along the column length showing supersaturation with respect to dolomite (\ce{CaMg(CO3)2}), magnesite (\ce{MgCO3}), and huntite (\ce{Mg3Ca(CO3)4}) using medium dissolution rate (k = \SI{1.0e-9}{mol.m^{-2}.s^{-1}}). (d) High-resolution images of isolated carbonate precipitate bodies on dark basaltic glass substrate, with scale bars indicating millimeter-scale accumulations, demonstrating the nucleation-controlled, stochastic nature of mineralization rather than uniform growth-dominated precipitation.}
    \label{fig:column_reactor}
\end{figure}

Figure~\ref{fig:column_reactor} presents the reactive transport experiment at \SI{0.005}{mL/min}. The column reactor configuration, consisting of an initial \SI{4}{cm} section packed with crushed calcium carbonate followed by \SI{36}{cm} of basaltic glass, is illustrated schematically alongside a photograph of the experimental setup (Fig.~\ref{fig:column_reactor}a). The calcium carbonate section served as a rapid source of dissolved \ce{Ca^{2+}} ions upon contact with the acidified seawater, while the basaltic glass section provided the primary reactive substrate for mineral precipitation.

Temporal monitoring at 7 and 21 days, along with end-state observations at 30 days, revealed surprising patterns of carbonate mineral accumulation (Fig.~\ref{fig:column_reactor}b). White crystalline patches, identified as carbonate precipitates, appeared randomly distributed along the entire length of the basalt-packed section. Contrary to expectations from deterministic reactive transport models, which predicted maximum precipitation near the calcite-basalt transition zone where \ce{Ca^{2+}} concentrations would be highest, a number of larger accumulations occurred in the second half of the column. This counterintuitive spatial distribution suggests that factors beyond simple concentration gradients govern precipitation patterns in the performed experiments.

The progressive pH increase along the column, from inlet pH 5.4--5.5 through calcite buffering to outlet pH 6.58--7.25, may impact precipitation patterns through pH-dependent carbonate speciation and nucleation kinetics. At lower pH, carbonate exists predominantly as \ce{H2CO3} and \ce{HCO3^-} rather than \ce{CO3^{2-}} required for calcite precipitation, while \ce{H^+} ions kinetically inhibit nucleation. Higher pH increases \ce{CO3^{2-}} availability and reduces nucleation energy barriers, potentially accelerating nucleation rates. This can explain enhanced precipitation in latter column sections despite lower \ce{Ca^{2+}} concentrations: pH-controlled nucleation kinetics dominate over concentration-driven thermodynamics, though stochastic nucleation dynamics prevent deterministic spatial prediction.

The observed spatial distribution additionally reflects probabilistic nucleation dynamics \cite{nooraiepour2021probabilistic, nooraiepour2021SciRepprobabilistic,masoudi2022effect}. Once initial carbonate precipitation occurs, subsequent nucleation may be significantly enhanced on these newly formed carbonate surfaces, a phenomenon known as precipitation on a secondary substrate, where carbonate-carbonate interfacial energies are lower than those of carbonate-silicate interfaces. This mechanism could also contribute to the isolated, growing accumulations at initially stochastic nucleation sites rather than uniformly distributed precipitation.

High-resolution imaging of selected precipitate bodies revealed their morphology and size distribution (Fig.~\ref{fig:column_reactor}d). Individual carbonate accumulations ranged from sub-millimeter to several millimeters in characteristic dimension, forming as discrete, isolated pockets rather than continuous coatings or uniformly dispersed aggregates. The contrast between white carbonate minerals and the dark basaltic glass substrate facilitated visual identification and quantification of precipitation zones. Notably, large precipitate bodies were spatially separated, with extensive regions of the column showing minimal visible mineralization, the dissolution of basaltic glass, and continuous \ce{Ca^{2+}} supply.

These observations collectively indicate that carbonate precipitation under the experimental conditions is predominantly controlled by nucleation kinetics rather than crystal growth rates. Once a stable nucleus forms on the basaltic glass surface, subsequent growth proceeds readily due to the continuous supply of supersaturated fluid. However, nucleation events themselves appear to be stochastic, occurring randomly along the column rather than in predictable locations based solely on supersaturation gradients. This nucleation-limited behavior has significant implications for predicting mineralization efficiency and spatial distribution in field-scale \ce{CO2} storage operations, as it introduces inherent uncertainty that cannot be captured by deterministic models alone.

The acidity of the injected \ce{CO2}-charged seawater was monitored routinely at the inlet and maintained at pH = 5.4--5.5 throughout the 30-day experimental period. Upon entering the column, this acidified fluid first contacted the crushed calcium carbonate section, where rapid buffering occurred. Aqueous geochemical calculations using PHREEQC v3 indicate that equilibration with calcite raised the pH to approximately 6.0, accompanied by substantial dissolution of \ce{CaCO3} and corresponding increase in dissolved \ce{Ca^{2+}} and carbonate species concentrations. This pre-buffered fluid then entered the basaltic glass section, where further buffering occurred through basaltic glass dissolution, releasing additional divalent cations (\ce{Ca^{2+}}, \ce{Mg^{2+}}, \ce{Fe^{2+}}) and progressively increasing the pH.

The measured pH of the outlet effluent exhibited a systematic decline during the experiments, decreasing from 7.25 at day 1 to 6.58 at day 30. This temporal pH evolution can be attributed to two competing processes: (1) decreasing basalt dissolution rates as reactive surfaces become passivated by secondary mineral coatings or as the most reactive glass components are preferentially leached, and (2) progressive precipitation of carbonate minerals within the column, which consumes alkalinity and shifts the carbonate equilibrium. Process (1) involves passivation by non-carbonate secondary phases, or clays coating reactive basalt surfaces and suppressing further dissolution, whereas process (2) specifically refers to the precipitation of carbonate minerals that directly consume \ce{CO3^2-} and \ce{HCO3^-}, thereby reducing solution alkalinity. The observed pH decline is consistent with the system transitioning from an initial disequilibrium toward a quasi-steady state, where dissolution and precipitation rates approach balance.

One-dimensional reactive transport simulations were conducted to evaluate fluid geochemical evolution along the column and to constrain basalt dissolution rates (Fig.~\ref{fig:column_reactor}c). Due to uncertainties in both reactive surface area and intrinsic dissolution rate constants for the basaltic glass, three scenarios spanning two orders of magnitude in glass dissolution rate were simulated: fast (k = \SI{1.0e-8}{mol.m^{-2}.s^{-1}}), medium (k = \SI{1.0e-9}{mol.m^{-2}.s^{-1}}), and slow (k = \SI{1.0e-10}{mol.m^{-2}.s^{-1}}) (Fig.~\ref{fig:column_reactor}c, top panel).

The simulations reveal that calcite supersaturation at the column outlet varies dramatically depending on the assumed dissolution rate, ranging from near-equilibrium conditions (saturation index $\sim$0) in the slow dissolution scenario to extreme supersaturation (saturation index $>$3, corresponding to $>$1000$\times$ supersaturation) in the fast dissolution case. Comparison of simulated outlet pH values (6.0, 6.5, and 9.5 for fast, medium, and slow dissolution scenarios, respectively) with the experimentally measured range (6.58--7.25) suggests that the actual basalt dissolution rate is comparable to or slightly higher than the medium case (k $\approx$ \SI{1.0e-9}{mol.m^{-2}.s^{-1}}). This rate estimate is consistent with values extrapolated from the temperature-dependent dissolution kinetics of basaltic glass across a broad pH range reported by \cite{oelkers2001mechanism,gislason2003mechanism}.

Thermodynamic speciation calculations indicate that multiple carbonate mineral phases are supersaturated along the column (Fig.~\ref{fig:column_reactor}c, bottom panel). Anhydrous magnesian carbonates including dolomite (\ce{CaMg(CO3)2}), magnesite (\ce{MgCO3}), and huntite (\ce{Mg3Ca(CO3)4}) are all thermodynamically stable and supersaturated, whereas hydrous magnesium carbonate phases such as nesquehonite (\ce{MgCO3}$\cdot$\ce{3H2O}) remain undersaturated under the experimental conditions. Calcite (\ce{CaCO3}) exhibits the highest supersaturation and fastest precipitation kinetics, consistent with its identification as the dominant precipitate phase in post-experiment characterization.

The progressive increase in carbonate mineral supersaturation along the flow path, coupled with the observed stochastic nucleation behavior, suggests that mineralization in this basaltic system operates in a kinetically controlled regime where nucleation barriers, rather than thermodynamic driving forces, govern precipitation rates and spatial patterns. It indicates that enhancing nucleation site density through surface roughening, seeding with carbonate particles (availability of secondary substrate), or controlled pH/alkalinity cycling may be more effective than simply increasing supersaturation levels for accelerating mineralization rates.

\subsection{Pore-scale mineralization and secondary phase formation}

\begin{figure}[h!]
    \centering
    \includegraphics[width=0.8\textwidth]{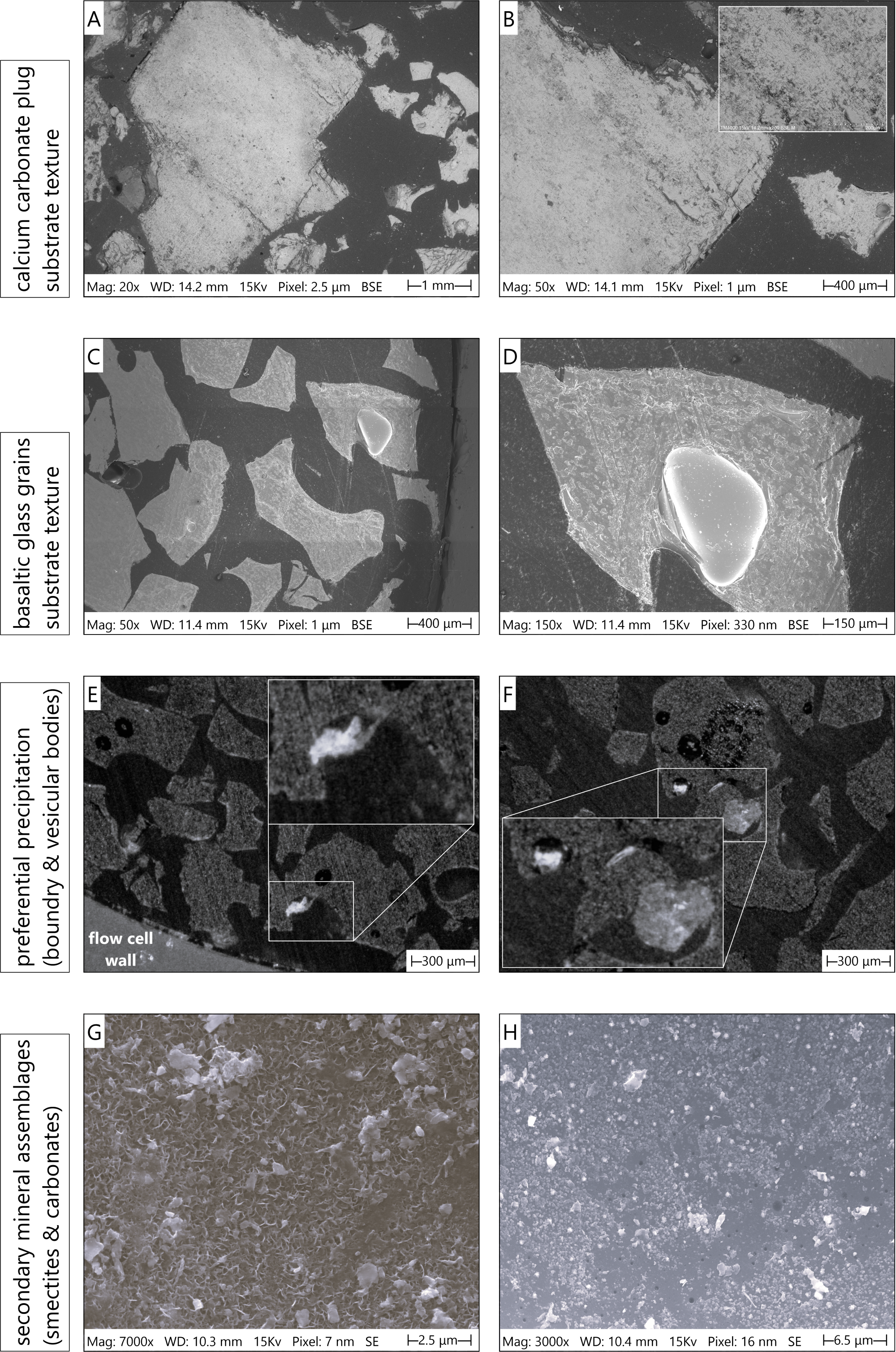}
        \caption{Pore-scale characterization of post-experiment substrates revealing dissolution features, spatial precipitation patterns, and secondary mineral assemblages. (a--b) Surface texture of calcium carbonate plug grains after reactive flow, showing dissolution features and surface roughening from acidified seawater interaction. (c--d) Basaltic glass grain surfaces, exhibiting characteristic dissolution morphology with enhanced surface roughness. (e) Preferential carbonate precipitation in peripheral regions adjacent to the flow cell wall, attributed to reduced advection velocity in the boundary layer and enhanced fluid retention on water-wet glass surfaces. (f) Enhanced carbonate mineralization within natural cavities and vesicular features of basaltic glass grains. (g) SEM micrograph showing pervasive smectite clay formation with characteristic texture on basaltic glass surfaces. (h) Early-stage carbonate formation, characterized by crystallites and amorphous calcium-rich precipitates, on basaltic glass surfaces.}
    \label{fig:pore_scale}
\end{figure}

Post-experiment characterization of basaltic glass substrates revealed complex mineralization patterns and secondary phase assemblages that deviate from ideal thermodynamic-geochemical predictions, highlighting the critical role of kinetic limitations, spatial heterogeneity, and competing reactions in controlling carbon mineralization efficiency. Detailed pore-scale analysis identified four principal phenomena that collectively governed reactive transport and mineral precipitation under flow-through conditions: (1) smectite clay formation, (2) preferential carbonate precipitation in peripheral regions of the flow cell, (3) early-stage calcium-enriched surface phases, and (4) enhanced mineralization within cavities and vesicular features of the basaltic substrate (Fig.~\ref{fig:pore_scale}).

\subsubsection{Substrate dissolution and surface modification}

SEM examination of post-experiment substrates revealed distinct dissolution-induced surface modifications on both calcium carbonate plug grains and basaltic glass particles (Fig.~\ref{fig:pore_scale}a--d). The calcium carbonate grains, initially included in the column inlet section to provide dissolved \ce{Ca^{2+}} through reaction with acidified seawater, exhibited pronounced surface roughening and the development of dissolution features including etch pits and surface relief patterns (Fig.~\ref{fig:pore_scale}a,b). These morphological changes reflect the dissolution regime imposed by the acidity of the injected fluid (pH 5.4--5.5), which liberated calcium ions into solution.

Basaltic glass grain surfaces similarly displayed characteristic dissolution morphology with significantly enhanced surface roughness compared to the pristine substrate (Fig.~\ref{fig:pore_scale}c,d). The dissolution process preferentially attacked structurally weak zones and created micro-scale surface irregularities that subsequently served as potential nucleation sites for secondary mineral precipitation. The textural evolution of basaltic glass surfaces during reactive flow demonstrates the dynamic nature of fluid-rock interactions, where simultaneous dissolution of primary phases and precipitation of secondary phases continuously modify the reactive surface area and properties available for subsequent mineralization reactions.

\subsubsection{Spatial heterogeneity in carbonate precipitation}

Detailed examination of post-experiment substrates revealed pronounced spatial heterogeneity in carbonate distribution, with three distinct localization patterns: preferential precipitation in peripheral regions near and adjacent to the flow cell wall, enhanced mineralization within cavities and vesicular features, in addition to the random distribution of isolated large crystals throughout the column interior (Fig.~\ref{fig:pore_scale}e,f). These patterns could reflect the complex interplay between hydrodynamics, surface properties, and nucleation kinetics in controlling the spatial distribution of mineralization.

Carbonate accumulation in peripheral regions near the glass tube wall (Fig.~\ref{fig:pore_scale}e) can be attributed to reduced advection velocities in the boundary layer, where viscous drag creates a low-velocity zone extending several grain diameters from the wall. This velocity reduction increases local fluid residence time, allowing progressive accumulation of dissolved cations and sustained supersaturation with respect to carbonate minerals. Additionally, the water-wet properties of the borosilicate glass tube may promote preferential fluid retention in wall-adjacent regions, further enhancing residence time effects. The hydrophilic glass surface facilitates the formation of a thin aqueous film, which experiences minimal advective transport, thereby creating conditions conducive to diffusion-controlled mineralization.

The vesicular and cavity-rich nature of basaltic glass grains provided preferential nucleation sites for carbonate precipitation (Fig.~\ref{fig:pore_scale}f). Surface irregularities, including natural cavities, vesicles, and conchoidal fracture features, offer geometrically favorable sites for heterogeneous nucleation by reducing the critical nucleus size through decreased interfacial energy. Within these micro-environments, fluid stagnation zones develop where advective transport is minimal, allowing diffusion-dominated conditions that favor progressive supersaturation buildup.

Significantly, the widespread occurrence of elevated calcium concentrations across basaltic glass surfaces (Fig.~\ref{fig:pore_scale}h), relative to the limited number of macroscopic carbonate precipitate patches, underscores the dominance of nucleation kinetics over growth kinetics in controlling overall mineralization rates. While thermodynamic conditions favorable for carbonate precipitation existed throughout the system, only specific locations—characterized by optimal combinations of supersaturation, residence time, and surface properties—progressed from initial calcium enrichment to visible crystal formation. This observation reinforces the stochastic, nucleation-limited nature of carbonate mineralization documented in the column-scale experiments.

\subsubsection{Secondary mineral assemblages: carbonates and clays}

The dissolution of basaltic glass released substantial quantities of divalent cations (\ce{Ca^{2+}}, \ce{Mg^{2+}}, \ce{Fe^{2+}}) and silica into solution, creating conditions for both carbonate and phyllosilicate precipitation. Post-reaction SEM characterization revealed two distinct secondary mineral assemblages: clay minerals exhibiting characteristic sheet-like morphology consistent with smectite group phyllosilicates (Fig.~\ref{fig:pore_scale}g), and early-stage calcium-enriched phases representing incipient carbonate formation (Fig.~\ref{fig:pore_scale}h). The smectite phases displayed the honeycomb-like texture typical of dioctahedral smectites, forming patchy (non-uniform and discontinuous distribution) coatings on basaltic glass surfaces. The presence of smectite may create activation energy barriers that kinetically inhibit carbonate nucleation on coated surfaces, analogous to the smectite-induced inhibition previously suggested for basaltic systems \cite{hellevang2017experimental}. This could potentially contribute to the spatially limited carbonate precipitation despite widespread supersaturation throughout the column.

The formation of smectite clays under our experimental conditions (80\,\textdegree{}C, pH 6.58--7.25) is consistent with previous studies of basalt alteration in \ce{CO2}-charged aqueous systems \cite{hellevang2017experimental}. Thermodynamic calculations indicate that smectite stability is favored at circumneutral pH and temperatures below 100\,\textdegree{}C, conditions where the \ce{Mg^{2+}}/\ce{H^{+}} activity ratio promotes phyllosilicate formation over carbonate precipitation. The pervasive nature of smectite coatings on basaltic glass surfaces suggests that clay formation represents a competing sink for divalent cations that would otherwise contribute to carbonate mineralization.

In contrast to the predicted formation of magnesian and ferroan carbonates (magnesite, siderite, ankerite) suggested by equilibrium thermodynamic modeling, the experimentally observed carbonate phases were predominantly calcium-rich. High-resolution EDS mapping revealed elevated calcium concentrations across extensive surface areas (Fig.~\ref{fig:pore_scale}h), indicating widespread nucleation of calcium carbonate precursors, though only a limited number of these nucleation sites developed into macroscopic crystalline precipitates. This observation suggests that while thermodynamic driving forces for carbonate precipitation exist throughout the system, kinetic barriers to rapid growth limit the spatial extent of actual mineralization.

Three mechanistic explanations may account for the absence of \ce{Mg-Fe-Ca} carbonates, despite their thermodynamic stability and supersaturation in the reacted fluids: (1) preferential sequestration of \ce{Mg^{2+}} and \ce{Fe^{2+}} by smectite formation, depleting solution concentrations below the threshold required for mixed carbonate nucleation; (2) kinetic inhibition of carbonate nucleation due to surface passivation by clay (smectite) coatings, which create activation energy barriers for heterogeneous nucleation; and (3) slow-moving crystal growth kinetics resulting from unfavorable aqueous \ce{Me^{2+}}/\ce{CO3^{2-}} activity ratios that extend nucleation induction times beyond experimental timescales. This has collective implications for reactive transport modeling. Kinetic frameworks based solely on transition state theory (TST)-derived rate laws, which do not explicitly account for nucleation barriers or the influence of aqueous activity ratios on precipitation kinetics, may overestimate carbonatization rates by several orders of magnitude \cite{hellevang2017experimental}. Accurate prediction of \ce{CO2} mineralization efficiency in basaltic systems requires incorporation of nucleation-based kinetic models that capture the stochastic nature of carbonate formation as well as the effect of solution stoichiometry \cite{hellevang2016statistical}.

The preferential formation of calcium carbonate over magnesium phases reflects differences in precipitation kinetics. Calcite exhibits faster nucleation rates and lower activation energies compared to magnesite, particularly at temperatures below 100\,\textdegree{}C \cite{pokrovsky2009effect}. The higher hydration energy of \ce{Mg^{2+}} relative to \ce{Ca^{2+}} imposes a substantial kinetic barrier to magnesite precipitation, requiring either elevated temperatures ($>$150\,\textdegree{}C) or extended reaction times to achieve significant magnesian carbonate formation. These kinetic limitations are exacerbated in flow-through systems where residence times are constrained, further favoring rapid calcite precipitation over slower-forming magnesian phases.

\subsection{Pore space characterization of vesicular basalts}

\begin{figure}[h!]
    \centering
    \includegraphics[width=1.1\textwidth]{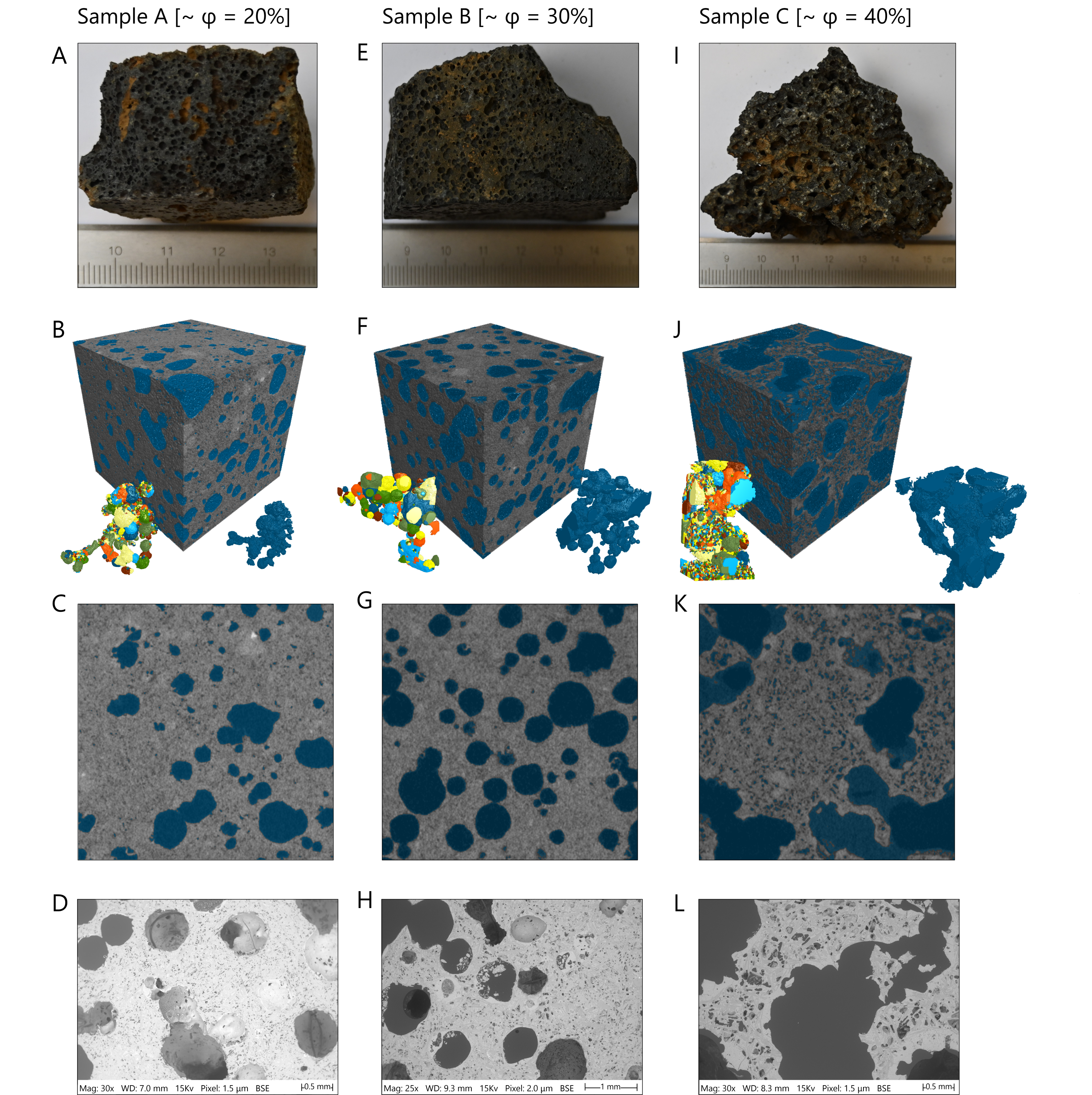}
    \caption{Multi-scale characterization of three Icelandic basalt facies representing the spectrum of vesicularity in basaltic reservoirs. Columns from left to right: Sample A (dense flow-interior facies) (A-D), Sample B (transitional facies)(E-H), Sample C (highly vesicular flow-top facies) (I-L). {First row:} Visible light photography showing macroscale textural features and vesicle distributions. {Second row:} Three-dimensional micro-CT reconstructions with segmented pore space (blue), the largest connected-component pore network, and corresponding extracted individual pore bodies derived from PNM. {Third row:} Vertical profiles through segmented volumes illustrating spatial distribution and connectivity of vesicular networks. {Fourth row:} High-resolution SEM micrographs characterizing vesicle wall thickness, inter-vesicle connectivity, and matrix microporosity. Note the progressively increasing matrix microporosity from Sample A through C, which contributes to overall network connectivity beyond the macro-vesicular structure alone and helps explain the expected superior hydraulic conductivity of Sample C relative to Sample B.}
    \label{fig:basalt_facies}
\end{figure}

Three distinct facies of Icelandic basalts, representing the spectrum of vesicularity expected in basaltic reservoirs, were characterized using integrated multi-scale imaging including visible light photography, high-resolution X-ray micro-computed tomography (micro-CT), and scanning electron microscopy (SEM) (Fig.~\ref{fig:basalt_facies}). These samples encompass: (1) dense flow-interior facies (Sample A) with limited vesicle connectivity, (2) transitional facies (Sample B) with intermediate connectivity, and (3) highly vesicular flow-top facies (Sample C) with well-developed pore networks. This suite provides systematic insight into how vesicle connectivity and pore network topology control reactive transport behavior in basaltic \ce{CO2} storage reservoirs.

Macro-scale photographic analysis reveals distinct textural characteristics across the three facies (Fig.~\ref{fig:basalt_facies}, first row). Sample C displays abundant vesicles ranging from millimeter to centimeter scale, with elongated geometries reflecting flow-induced deformation during magma emplacement. Sample B exhibits moderate vesicularity with a bimodal size distribution, comprising small (1--2\,mm) spherical vesicles and larger (>5\,mm) irregular voids. Sample A appears substantially denser and more massive, with scattered, isolated vesicles discernible only upon close examination of macroscale features throughout the specimen.

Three-dimensional micro-CT reconstructions (Fig.~\ref{fig:basalt_facies}, second row) coupled with vertical profiles through the segmented volumes (Fig.~\ref{fig:basalt_facies}, third row) quantify fundamental topological differences between the basaltic facies. For each sample, two specimens were imaged and analyzed to capture inherent heterogeneities. Segmented porosity analysis reveals total porosities of 18.3--21.0\% ($\pm$2\%) for Sample A, 30.5--31.3\% ($\pm$3\%) for Sample B, and 41.4--41.5\% ($\pm$5\%) for Sample C. Region-of-interest (ROI) size-dependent variations in measured porosity were observed, particularly for Sample C, where inclusion of large coalesced vesicular bodies introduced notable pore volume increases, resulting in 1--5\% variability across the sample suite based on the selected ROI size. 

The fraction of porosity contributing to the largest connected-component pore network (visualized alongside the 3D segmented volumes in the second row) represents a critical parameter controlling bulk permeability. It differs dramatically across samples: only 1.3--3.8\% distributed over multiple disconnected ROIs in Sample A, 12.1--13.3\% in Sample B, and 31.7--32.2\% in Sample C. This contrast indicates that Sample A contains predominantly isolated, non-effective porosity, while Sample C exhibits nearly complete pore connectivity throughout the vesicle network. The spatial distribution and connectivity patterns are clearly visualized in the vertical profiles presented in the third row, where Sample C shows several continuous, connected pathways, Sample B reveals partially connected networks with interruptions, and Sample A displays smaller, predominantly isolated pore clusters.

High-resolution SEM imaging provides micrometer-scale characterization of vesicle morphology, surface texture, and inter-vesicle connectivity (Fig.~\ref{fig:basalt_facies}, fourth row). In Sample C, vesicle walls are thin and frequently breached, creating direct hydraulic connections between adjacent voids through inter-vesicle windows ranging from tens of micrometers to millimeters in diameter. These breached connections enable efficient fluid transport throughout the pore network. Sample A reveals predominantly isolated vesicles with thick, intact walls and rare inter-vesicle connections. Surface texture analysis reveals significant matrix microporosity in Sample C, which progressively decreases in size and abundance from Sample C to Sample A, reflecting variations in cooling rate and degassing efficiency during lava emplacement.

Quantitative pore structure analysis reveals systematic trends correlating with bulk porosity. Mean pore diameter increases progressively: 0.25--1.4\,mm in Sample A, 0.6--1.5\,mm in Sample B, and 0.7--2.2\,mm in Sample C. Sample C exhibits elongated, interconnected vesicular structures forming extensive void spaces spanning several millimeters without narrow constrictions, whereas Sample A contains smaller, more isolated vesicular bodies. Measured pore throat diameters—the critical bottlenecks controlling permeability—range from 0.15--0.40\,mm in Sample A, 0.25--0.75\,mm in Sample B, and 0.3--1.0\,mm in Sample C. The narrow throats in Sample A, combined with poor connectivity, yield lower expected fluid permeability, while the wider, well-connected throats in Sample C enable highly permeable percolation pathways.

Pore coordination number distributions extracted from the 3D micro-CT datasets reveal dramatic contrasts in network architecture topology. Sample C exhibits coordination numbers ranging from 1 to 4 with a modal value of 2, indicating that most vesicles connect to two neighboring voids, creating predominantly serial or chain-like connectivity. Sample B shows similar coordination number distributions but with reduced overall connectivity due to lower vesicle abundance and incomplete coalescence. Sample A displays extremely low coordination, with the majority of vesicles either completely isolated (coordination number = 0) or connected to only a single neighbor (coordination number = 1), confirming the absence of percolating flow networks capable of sustaining effective fluid transport.

The integrated multi-scale characterization demonstrates systematic evolution of pore network architecture from dense flow interiors through transitional zones to highly vesicular flow tops. This progression reflects the interplay between vesicle nucleation density, bubble growth and coalescence during magma ascent and lateral flow, and cooling rate during solidification. The dramatic differences in connected porosity fraction (spanning from <4\% to >30\%) despite more modest variations in total porosity (18--41\%) underscore the critical importance of pore connectivity topology rather than simple porosity magnitude in controlling hydraulic properties and suitability for \ce{CO2} storage applications.

Quantitative pore network modeling (PNM) of the segmented micro-CT volumes provides detailed statistical characterization of individual pore bodies (vertices) and connecting throats (edges) across the three basalt facies (Table~\ref{tab:pnm_statistics}). PNM analysis extracts topological and geometric properties from the three-dimensional pore space, enabling systematic comparison of pore and throat size distributions, connectivity metrics, and network architecture parameters that control fluid transport.

\begin{table}[htbp]
\footnotesize
\centering
\caption{Pore network modeling statistics for vesicular basalt samples derived from micro-CT segmentation. Vertex (pore body) and edge (throat) scalar values represent distributions across the extracted pore networks. Values reported as mean ± standard deviation (indicating statistical spread) with 5th--95th percentile ranges in parentheses showing actual data bounds. Pore-specific surface area (SSA$_{\text{pore}}$) is the ratio of mean surface area to mean pore volume. Bulk-specific surface area (SSA$_{\text{bulk}}$) is SSA$_{\text{pore}} \times \phi$, using mid-point porosity for Samples A-C.}
\label{tab:pnm_statistics}
\begin{tabular}{@{}llccc@{}}
\toprule
\textbf{Parameter} & \textbf{Unit} & \textbf{Sample A} & \textbf{Sample B} & \textbf{Sample C} \\
\midrule
\multicolumn{5}{l}{\textit{Vertex (Pore Body) Properties}} \\
\midrule
Connectivity & -- & 3.2 ± 2.1 & 3.2 ± 2.6 & 5.6 ± 5.1 \\
 & & (1.0--6.8) & (1.0--7.8) & (1.0--12.5) \\
\addlinespace[0.1cm]
Volume & mm³ & 0.01 ± 0.04 & 0.16 ± 0.48 & 0.06 ± 0.12 \\
 & & (0.0001--0.0001) & (0.0001--0.51) & (0.00004--0.00004) \\
\addlinespace[0.1cm]
Surface area & mm² & 0.09 ± 0.26 & 0.74 ± 1.52 & 0.17 ± 0.49 \\
 & & (0.02--0.17) & (0.004--2.95) & (0.02--0.02) \\
\addlinespace[0.1cm]
{Pore-specific SSA} & {mm}$^{{-1}}$ & {9.00} & {4.63} & {2.83} \\
\addlinespace[0.1cm]
{Bulk-specific SSA} & {mm}$^{{-1}}$ & {1.76} & {1.43} & {1.17} \\
\addlinespace[0.1cm]
Inscribed diameter & mm & 0.09 ± 0.06 & 0.24 ± 0.24 & 0.09 ± 0.06 \\
 & & (0.05--0.16) & (0.03--0.73) & (0.05--0.15) \\
\addlinespace[0.1cm]
Equivalent diameter & mm & 0.15 ± 0.09 & 0.35 ± 0.33 & 0.14 ± 0.10 \\
 & & (0.07--0.28) & (0.06--0.97) & (0.04--0.26) \\
\midrule
\multicolumn{5}{l}{\textit{Edge (Throat) Properties}} \\
\midrule
Equivalent diameter & mm & 0.08 ± 0.06 & 0.21 ± 0.24 & 0.08 ± 0.07 \\
 & & (0.01--0.17) & (0.01--0.72) & (0.01--0.17) \\
\addlinespace[0.1cm]
Total length & mm & 0.24 ± 0.13 & 0.64 ± 0.43 & 0.28 ± 0.21 \\
 & & (0.11--0.50) & (0.17--1.49) & (0.11--0.71) \\
\addlinespace[0.1cm]
Cross-sectional area & mm² & 0.01 ± 0.02 & 0.09 ± 0.23 & 0.02 ± 0.04 \\
 & & (0.0001--0.02) & (0.0001--0.42) & (0.0001--0.0001) \\
\bottomrule
\end{tabular}
{\footnotesize \parbox{\linewidth}{\raggedright Connectivity represents the number of throats connected to each pore body in the entire volume. Inscribed diameter is the diameter of the largest sphere that fits within the pore. Equivalent diameter is the diameter of a sphere with equal volume. Cross-sectional area refers to the throat constriction area. Sample A connectivity values represent local clustering within isolated pore groups rather than network-scale connectivity.}}
\end{table}

Vertex (pore body) analysis reveals systematic variations in pore geometry and connectivity across the three facies. Sample B exhibits the largest mean pore volumes (0.16 ± 0.48\,mm³) and surface areas (0.74 ± 1.52\,mm²), with equivalent diameters reaching 0.35 ± 0.33\,mm and 95th percentile values approaching 1\,mm. These substantial pore dimensions reflect the transitional nature of this facies, where moderate vesicularity combines with partial coalescence to create intermediate-scale void spaces. Sample C, despite higher total porosity (41\%), displays smaller mean pore volumes (0.06 ± 0.12\,mm³) due to the significant share of porosity identified in the basalt texture itself, and equivalent diameters (0.14 ± 0.10\,mm), indicating that the elevated porosity derives from abundant small-to-intermediate pore bodies (in the basalt framework and as vesicles) besides the many large coalesced voids. Sample A exhibits the smallest pore volumes (0.01 ± 0.04\,mm³) and equivalent diameters (0.15 ± 0.09\,mm), consistent with its dense, poorly vesicular character.

The connectivity metric, which represents the number of throats connected to each pore body, provides critical insights into the network topology. Sample C exhibits the highest mean connectivity (5.6 ± 5.1), with 95th percentile values reaching 12.5, indicating a well-developed pore network where individual vesicles connect to multiple neighbors. Sample B displays intermediate connectivity (3.2 ± 2.6), reflecting partial network development. Sample A exhibits a connectivity value of 3.2 ± 2.1, which paradoxically appears comparable to Sample B despite dramatically lower bulk connected porosity (1.3--3.8\% versus 12.1--13.3\%). This apparent contradiction reflects the nature of PNM analysis in poorly connected systems: the connectivity values represent local coordination within small, isolated pore clusters rather than network-scale connectivity. Sample A comprises numerous isolated vesicle groups, each internally showing modest connectivity, but with minimal inter-cluster connections preventing the formation of percolating pathways. The large standard deviations across all samples (approaching or exceeding mean values) underscore the pronounced heterogeneity characteristic of vesicular pore networks.

Edge (throat) analysis quantifies the critical bottlenecks controlling permeability. Sample B exhibits substantially larger mean throat diameters (0.21 ± 0.24\,mm) compared to Samples C and A (both 0.08 ± 0.06--0.07\,mm), with 95th percentile values reaching 0.72\,mm. These wide throats, combined with the intermediate total porosity and connectivity, position Sample B as having favorable hydraulic properties despite not exhibiting the highest vesicularity. The throat lengths in Sample B (mean 0.64 ± 0.43\,mm, 95th percentile 1.49\,mm) exceed those in Samples C (0.28 ± 0.21\,mm) and A (0.24 ± 0.13\,mm), reflecting the elongated, tubular geometry of inter-vesicle connections in the transitional facies. Throat cross-sectional areas follow similar trends, with Sample B (0.09 ± 0.23\,mm²) having larger values than Samples C and A (0.02 ± 0.04 and 0.01 ± 0.02\,mm², respectively).

Pore-specific surface area (SSA$_{\text{pore}}$), decreases systematically from {9.0\,mm$^{-1}$} (dense flow-interior facies, Sample A) to {4.6\,mm$^{-1}$} (transitional facies, Sample B) and {2.8\,mm$^{-1}$} (highly vesicular flow-top facies, Sample C), reflecting a progressive increase in mean pore size and vesicle interconnectivity across the vesicularity spectrum (mean porosity from 19.5\% to 41.5\%). This inverse relationship indicates a petrophysical trend in basaltic rocks, where higher vesicularity favors larger, aperture-dominated pore networks, which in turn diminish the surface-to-volume ratio. Bulk SSA (SSA$_{\text{bulk}} =$ SSA$_{\text{pore}} \times \phi$), normalized to rock volume, follows suit at approximately {1.8 $\rightarrow$ 1.4 $\rightarrow$ 1.2\,mm$^{-1}$}.

The combination of pore and throat metrics explains the observed permeability hierarchy. Sample C achieves high permeability through abundant, well-connected small-to-intermediate pores despite modest individual throat diameters—the sheer number of parallel pathways compensates for individual constriction sizes. Sample B maintains intermediate-to-high permeability through fewer but substantially wider throats connecting larger pore bodies with a different architectural solution, achieving similar hydraulic conductivity. Sample A exhibits far lower permeability due to the combination of small pore volumes, narrow throats, and critically poor network-scale connectivity despite local clustering.

\subsection{Environmental implications for reactive flow engineering in vesicular basalt reservoirs}

\subsubsection{Vesicular pore architecture and precipitation-induced clogging}

Vesicular features in basaltic rocks form during magma solidification through a cascade of processes: volatile exsolution, bubble nucleation, growth, migration, and coalescence \cite{sparks1978gas,toramaru2022vesiculation,millett2024lava}. As magma ascends, the decrease in confining pressure reduces volatile solubility, driving supersaturation and triggering heterogeneous bubble nucleation on crystal surfaces or homogeneous nucleation within the melt \cite{gardner2023bubble}. These nascent bubbles grow through diffusive influx of volatiles from the surrounding melt and decompression-driven expansion \cite{gardner2023bubble,toramaru2022vesiculation}. During continued ascent and lateral flow, bubbles coalesce when inter-bubble films rupture, creating interconnected voids whose extent depends on ascent rate, bubble number density, and melt viscosity \cite{castro2012mechanisms,gardner2023bubble}. Upon solidification, this vesiculation history is preserved as a frozen record manifesting as vesicles ranging from submillimeter spherical pores to elongated, centimeter-scale interconnected structures that fundamentally control hydraulic properties.

A common misconception is that vesicles form only isolated, ineffective porosity. While isolated vesicles certainly exist within flow interiors and transitional zones \cite{millett2024lava,aubele1988vesicle,couves2016use,}, laboratory measurements and theoretical models demonstrate high connectivity above a percolation threshold of approximately 30\% porosity \cite{saar1999permeability,mueller2005permeability,bai2010permeability}. This critical threshold is commonly exceeded in basaltic flow tops \cite{cashman1997reevaluation,millett2024lava}, where porosities reach 40--50\% or higher, as confirmed by our Sample C characterization showing 41\% total porosity with 32\% connected porosity.

\begin{figure}[h!]
    \centering
    \includegraphics[width=\textwidth]{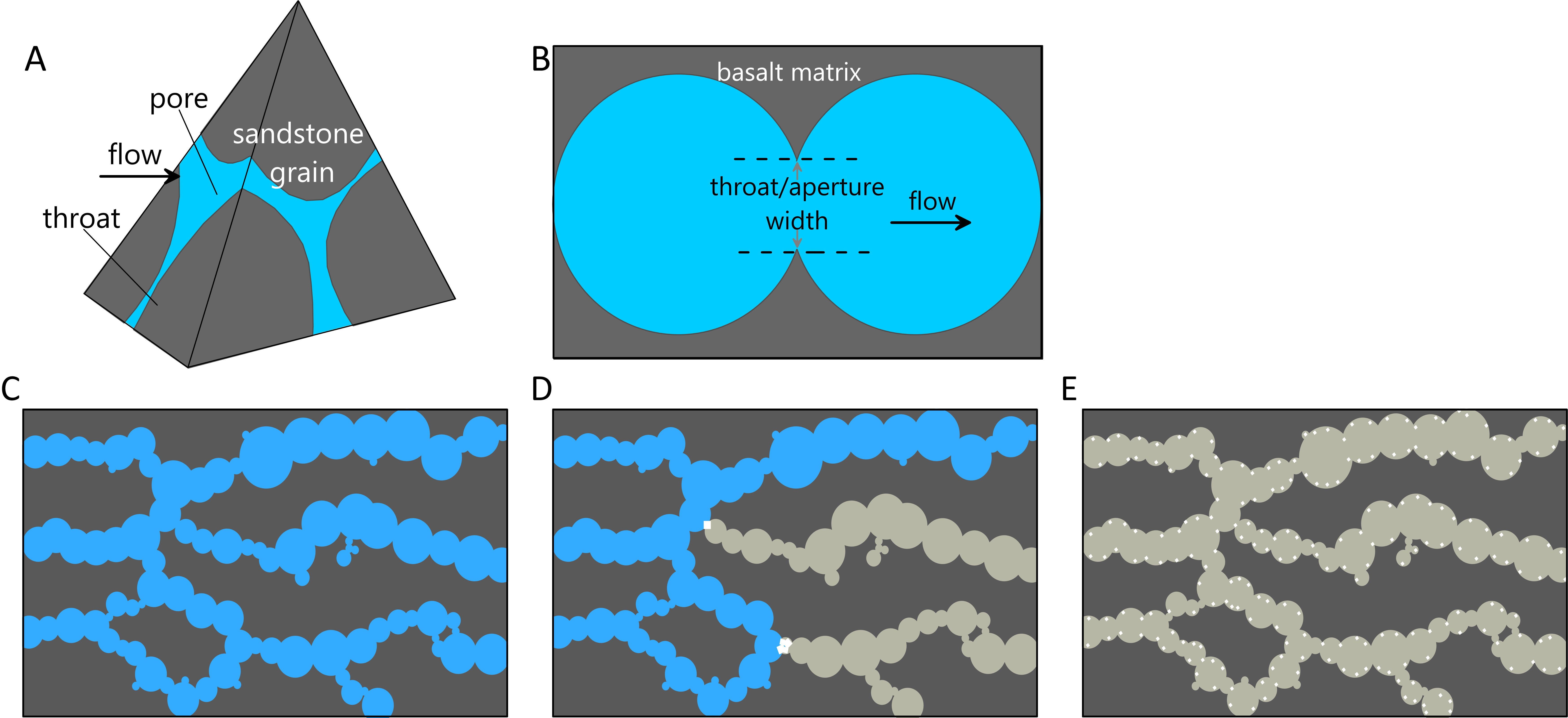}
    \caption{Pore network topology and precipitation-induced clogging scenarios in sandstone versus basalt reservoirs. {Top row:} Comparison of pore and throat architecture showing fundamental differences in network connectivity. (a) Representative sandstone pore network with high coordination number (4--6 connections per pore), creating multiple parallel flow pathways with substantial redundancy that maintains permeability even when individual throats become blocked. (b) Vesicular basalt pore network displaying low coordination number (predominantly 2 connections per pore), creating serial or chain-like connectivity vulnerable to catastrophic permeability loss if critical throats are occluded. Despite larger throat diameters in basalts compared to sandstones, the low-redundancy topology renders basaltic networks more susceptible to precipitation-induced impairment. {Bottom row:} Precipitation scenarios and their differential impacts on permeability in basaltic pore networks. (c) Schematic representation of percolation pathways in highly vesicular basalt reconstructed from micro-CT imaging, illustrating the tenuous, predominantly serial connectivity characteristic of vesicular networks. (d) Scenario 1: Large, isolated carbonate precipitates (white masses) forming within two main flow percolation pathways with varying degrees of adverse consequences. (e) Scenario 2: Numerous small, distributed carbonate precipitates (white patches) systematically impact multiple pore throats throughout the network. This scenario causes severe permeability degradation despite modest porosity loss, as extensively studied in \cite{masoudi2024mineral}.}
    \label{fig:clogging_scenarios}
\end{figure}

Permeability evolution with porosity in vesicular basalts follows non-linear relationships that are fundamentally different from those in granular sedimentary rocks, driven primarily by differences in pore network topology. Pore coordination number, which is the number of throats connecting each pore body to its neighbors, serves as the critical metric quantifying these topological differences. As illustrated in Figure~\ref{fig:clogging_scenarios}a, reservoir-quality sandstones exhibit coordination numbers ranging from 4 to 6, reflecting geometrically regular packing of approximately spherical grains. This high coordination creates multiple parallel flow pathways, imparting substantial redundancy such that blockage of individual pore throats has a limited impact on bulk permeability, since fluids redistribute through alternative connections.

In contrast, vesicular basalts exhibit dramatically lower coordination numbers (1--3, with modal value of 2), reflecting their fundamentally different origin through bubble coalescence rather than grain packing (Fig.~\ref{fig:clogging_scenarios}b). This creates serial or chain-like connectivity rather than three-dimensional networks, where many vesicles connect to only one or two neighbors, establishing low-redundancy serial flow architecture. Notably, vesicular basalts often display larger pore apertures (throat diameters of 0.3--1.0\,mm in Sample C) compared to sandstone matrix pores (typically <0.05-0.1\,mm), yet the low coordination number renders them more vulnerable to permeability impairment despite these wider individual connections.

The difference in coordination number between sandstones (4--6) and basalts (1--3) fundamentally alters reactive transport vulnerability to mineral precipitation. In high-coordination sandstones, precipitation must occlude a substantial fraction of pore throats, approximately 75\% based on percolation theory, before causing a significant reduction in permeability, as fluids continuously redistribute through alternative pathways. In low-coordination basalts, however, occlusion of even a single critical throat can sever connectivity across entire network portions, causing major local permeability loss. The permeability-porosity relationship becomes highly non-linear with threshold behavior: modest precipitation volumes can trigger disproportionate permeability reductions once critical bottlenecks seal. For coordination-2 systems, percolation theory predicts bond percolation thresholds near 50\%, meaning occlusion of approximately half the throats causes complete loss of connected pathways—substantially lower resilience than coordination-6 sandstones requiring $\sim$75\% blockage.

The pronounced spatial heterogeneity characteristic of vesicular basalts further exacerbates this topological vulnerability. Vesicle size distributions can span two to three orders of magnitude—from sub-millimeter spherical voids to centimeter-scale elongated cavities—with the smallest inter-vesicle connections representing critical choke points controlling bulk permeability. Preferential nucleation and growth of secondary minerals in these narrow throats, driven by locally elevated supersaturation and favorable surface properties for heterogeneous nucleation, amplifies the tendency toward early permeability impairment.

The interaction between mineral precipitation and fluid flow operates as a coupled, self-reinforcing feedback system with fundamentally different dynamics than in sandstone reservoirs. Under initial uniform injection conditions in unreacted basaltic samples, fluid velocity exhibits pronounced spatial heterogeneity reflecting the irregular vesicle geometry and variable throat dimensions. High velocities concentrate in large vesicles and wide connecting throats, the primary flow highways through the network, while stagnant or slowly recirculating fluid occupies smaller vesicles and dead-end pores branching from the primary pathways. This velocity heterogeneity establishes the initial template for reactive transport partitioning.

Carbonate precipitation initiates where local supersaturation reaches critical nucleation thresholds, which may not necessarily coincide with high-velocity flow paths. As documented in our column experiments, nucleation appears to be stochastic, dominated by surface properties, micro-scale heterogeneities, and local residence time, rather than simply tracking regions of highest bulk supersaturation. Once stable nuclei form, crystal growth proceeds at rates determined by the interplay of supersaturation, available growth surface area scaling with the mass of the secondary crystals, and mass transfer limitations. Critically, the growth of crystals progressively narrows pore throats, locally increasing fluid velocity and shear stress in constricted regions.

The probabilistic nature of precipitation location creates fundamentally different permeability evolution scenarios (Fig.~\ref{fig:clogging_scenarios}c--e). If large carbonate masses nucleate and grow within primary flow percolation pathways, the precipitates effectively occupy and disconnect flow highways (Fig.~\ref{fig:clogging_scenarios}d). These small yet chunky-enough precipitation volumes cause disproportionate permeability reduction by creating serial bottlenecks that restrict flow throughout the connected network.

Our recent pore-scale reactive transport simulations \cite{masoudi2024mineral} reveal a counterintuitive finding with significant implications for reservoir management: numerous small, distributed precipitates cause more severe permeability degradation than a few large, isolated accumulations (Fig.~\ref{fig:clogging_scenarios}e). While conventional intuition suggests that large carbonate patches represent the primary threat to injectivity, distributed fine-scale precipitation systematically reduces the effective diameter of multiple pore throats in parallel, eliminating the limited redundancy that might otherwise preserve permeability in low-coordination networks. This distributed clogging progressively converts the pore network from a poorly connected three-dimensional structure toward isolated, disconnected vesicle clusters, catastrophically reducing permeability despite very modest overall porosity loss. Low coordination number amplifies these precipitation-induced permeability reduction effects. In coordination-2 systems where each pore connects to only two neighbors, blockage of any single connection severs the entire linear chain, partitioning the network into isolated upstream and downstream segments with no alternative bypass pathways.

\subsubsection{Seawater versus freshwater for carbon mineralization}

The substitution of seawater for freshwater introduces geochemical constraints that alter mineralization pathways and efficiency. Seawater's elevated \ce{Mg^{2+}} ($\sim$53\,mM) and \ce{SO4^{2-}} ($\sim$28\,mM) concentrations create competing reactions that systematically reduce carbon mineralization as observed and also reported in literature \cite{voigt2021experimental,wolff2018flow,rosenbauer2012carbon}.

The high \ce{Mg^{2+}}/\ce{Ca^{2+}} ratio ($\sim$5:1) in seawater thermodynamically favors magnesian carbonates and aragonite. Aragonite tends to form elongated acicular crystals, making fibrous precipitates potentially having a relatively large reducing effect on permeability, while not altering porosity to any significant extent. The precipitation kinetics of magnesian carbonates, on the other hand, remain prohibitively slow below 100\,\textdegree{}C \cite{pokrovsky2009effect,saldi2009magnesite,}. Experimental evidence consistently demonstrates this kinetic barrier: Voigt et al. \cite{voigt2021experimental} achieved magnesite formation only at elevated pCO$_2$ (16\,bar) after 140 days, while Rosenbauer et al. \cite{rosenbauer2012carbon} observed ferroan magnesite requiring 100\,\textdegree{}C and months-long reaction times. At our experimental conditions, calcium carbonate (most probably aragonite) dominated over any magnesite (not observed in crystalline form) despite its appearance in thermodynamic-geochemical predictions.

Dissolved \ce{Mg^{2+}} simultaneously inhibits calcite precipitation through multiple mechanisms: surface adsorption reduces growth rates by up to 90\% \cite{davis2000role,liendo2022factors}, incorporation creates lattice strain \cite{meldrum2008controlling,davis2004morphological}, and competitive complexation reduces free \ce{Ca^{2+}} activity \cite{wolff2011dissolution}. Our flow-through experiments confirmed this inhibition, requiring order-of-magnitude residence time increases to achieve visible precipitation compared to typical freshwater systems.

Secondary mineral competition further diminishes mineralization efficiency. Our experiments revealed extensive smectite formation, corroborating observations by Voigt et al. \cite{voigt2021experimental} and Gysi and Stefánsson \cite{gysi2012co2,gysi2012experiments,} which documented Mg-saponite consuming available \ce{Mg^{2+}}. This phyllosilicate precipitation represents an irreversible cation sink while potentially reducing permeability through pore-throat occlusion, creating a dual impediment to effective storage.

Quantitative disparities between seawater and freshwater systems are striking: CarbFix achieved 95\% mineralization within two years using freshwater \cite{matter2016rapid}, while seawater experiments consistently yield <20\% over comparable timescales in batch \cite{voigt2021experimental,marieni2020experimental} and flow-through (current study) conditions. Wolff-Boenisch and Galeczka \cite{wolff2018flow} required artificial supersaturation to achieve carbonate precipitation in seawater at 90\,\textdegree{}C, conditions rarely sustainable in natural systems.

Despite these limitations, engineering strategies may partially overcome seawater's constraints. Weak acid co-injection can maintain a favorable pH while preventing excessive clay formation \cite{wolff2018flow,galeczka2014experimental}. Alternatively, hybrid approaches using initial freshwater slugs to establish carbonate precipitation zones followed by seawater injection may optimize both water usage and mineralization efficiency. Temperature optimization around 90-130\,\textdegree{}C balances enhanced kinetics against anhydrite precipitation risks \cite{voigt2021experimental}, though careful monitoring remains essential to prevent near-wellbore permeability loss.

Seawater serves as the primary fluid for offshore basaltic storage, aside from supercritical fluid injection, where containment is assured. Solubility trapping offers a reliable mechanism for intermediate-term storage security, enhancing \ce{CO2} sequestration in the dissolved phase while mineralization advances gradually in diffusion-dominated zones. The slower mineralization rates are mitigated by the extensive storage capacity of basaltic reservoirs, which feature dual-porosity systems comprising vesicular networks and fractured matrices. These systems sustain favorable injectivity despite partial mineral precipitation. Through effective pressure management and reservoir engineering, \ce{CO2} storage operations can accommodate prolonged storage operation timescales without sacrificing injection efficiency, enabling permanent carbon fixation over decades.

\section*{Conclusions}
This integrated experimental and numerical study reveals fundamental controls on \ce{CO2} mineralization in basaltic systems that challenge conventional reactive transport paradigms.

Carbonate precipitation under flow conditions is nucleation-controlled and stochastic rather than growth-controlled and deterministic. Isolated precipitate pockets form randomly along flow paths despite continuous supersaturation, demonstrating that thermodynamic predictions of supersaturation cannot forecast actual nucleation locations or timing. Residence time exerts primary control: an order-of-magnitude flow rate reduction (0.05 to 0.005 mL/min) proved necessary for visible carbonate formation, creating spatial partitioning between high-flux, low-mineralization flow highways and low-flux, high-mineralization matrix blocks.

Multi-scale characterization reveals that vesicular basalts exhibit coordination numbers (modal value = 2) three-fold lower than sandstones (modal value = 5), creating serial rather than parallel flow architecture. Connected porosity fractions (1.3--32.2\%) differ dramatically from total porosity (18--42\%), emphasizing that topology, not magnitude, controls permeability. This low-redundancy architecture renders basalts vulnerable to catastrophic permeability loss—percolation theory predicts connectivity loss at ~50\% throat blockage in coordination-2 basalts versus ~75\% in coordination-6 sandstones.

Secondary mineral assemblages comprise calcite-dominated carbonates and smectite clays, with magnesium carbonates absent despite thermodynamic favorability, reflecting kinetic limitations below 100°C. Smectite formation sequesters divalent cations, passivates reactive surfaces, and occludes pore throats, reducing mineralization efficiency but contributing to permanent \ce{CO2} sequestration through clay-associated trapping.

Seawater use complicates geochemistry, reduces predictability, and likely decreases mineralization rates compared to freshwater, though weak acid injection strategies leveraging solubility trapping may mitigate these limitations with proper engineering.

Successful basaltic \ce{CO2} storage requires: (1) probabilistic nucleation frameworks in reactive transport models; (2) stochastic reative transport modeling obtaining uncertainties in por-perm relations and flow impairment; (3) realistic pore network topologies from imaging data; (4) explicit treatment of competing clay reactions; (5) adaptive injection management preventing near-wellbore clogging; and (6) conservative rate estimates accounting for precipitation-permeability coupling. The low-coordination topology of vesicular basalts creates both opportunities—high initial permeability—and vulnerabilities—catastrophic permeability loss from modest precipitation—necessitating fundamentally different approaches than those used in conventional sandstone reservoir management.

\section*{Acknowledgments}
This work was supported by the “Understanding Coupled Mineral Dissolution and Precipitation in Reactive Subsurface Environments” project, funded by the Norwegian Centennial Chair (NOCC) as a transatlantic collaboration between the University of Oslo (Norway) and the University of Minnesota (USA).


\subsection*{Conflicts of Interest}
The authors declare no conflict of interest regarding the publication of this article.



\printbibliography

\end{document}